\documentclass[showpacs,amsmath,amssymb,prl,twocolumn]{revtex4-1}

\usepackage{graphicx}
\usepackage{dcolumn}
\usepackage{bm}

\begin{document}

\preprint{APS/123-QED}

\title{Polarization squeezing of light by single passage through an atomic vapor}

\author{S. Barreiro, P. Valente, H. Failache and A. Lezama}
\email{alezama@fing.edu.uy}
\affiliation{Instituto de F\'{\i}sica, Facultad de Ingenier\'{\i}a,
Universidad de la Rep\'{u}blica,\\ J. Herrera y Reissig 565, 11300
Montevideo, Uruguay}%

\date{\today}

\begin{abstract}
We have studied relative-intensity fluctuations for a variable set of orthogonal elliptic polarization components of a linearly polarized laser beam traversing a resonant $^{87}$Rb vapor cell. Significant polarization squeezing at the threshold level (-3dB) required for the implementation of several continuous variables quantum protocols was observed. The extreme simplicity of the setup, based on standard polarization components, makes it particularly convenient for quantum information applications.
\end{abstract}

\maketitle

In recent years, large attention has been given to the use of continuous variables for quantum information processing. A foreseen goal is the distribution of entanglement between distant nodes. For this, quantum correlated light beams are to interact with separate atomic systems in order to build quantum mechanical correlations between them \cite{DUAN01,BRAUNSTEIN05}.

A particular kind of quantum correlation between two light beams occurs when the intensity difference between them has fluctuations smaller than the standard quantum limit (SQL), that is smaller than the fluctuations of the intensity difference of two coherent states of the same intensity. The two beams are said to present relative-intensity squeezing (RIS). RIS has been generated through different nonlinear optics techniques. One of the most successful is parametric down conversion in a nonlinear $\chi^{(2)}$ crystal. Up to 9.7 dB RIS has been obtained with this method \cite{LAURAT05}. Such experiments require a relatively elaborate and expensive setup. The resulting light beams are spectrally broad and usually far detuned from convenient alkali atoms D  transitions. An alternative approach has considered the use of four-wave mixing in atomic samples \cite{MCCORMICK07,MCCORMICK08,GLORIEUX10,JASPERSE11}.

Reduced relative-intensity fluctuations have been observed between light beams of different frequency. However, RIS can also occur between orthogonal polarization components of a single light beam. In such case, the field is said to be polarization squeezed \cite{KOROLKOVA02,JOSSE03} and the noise reduction is described in terms of squeezing of the fluctuations of one of the Stokes operators:
\begin{eqnarray}
\nonumber S_{1}&=& a^{\dagger}_{x} a_{x}-a^{\dagger}_{y} a_{y},\\
\nonumber  S_{2}&=& a^{\dagger}_{x} a_{y}+a^{\dagger}_{y} a_{x},\\
\nonumber  S_{3}&=& i(a^{\dagger}_{y} a_{x}-a^{\dagger}_{x} a_{y})
\end{eqnarray}
Here $a_{x}, a_{y}$ are the field destruction operators for the orthogonal linear polarizations $\mathbf{x}$ and $\mathbf{y}$.

Polarization squeezing has been produced via propagation in optical fibers \cite{BOIVIN96,SILBERHORN01}, through the combination on a polarizing beam splitter of two quadrature squeezed light beams \cite{BOWEN02} and through the interaction of linearly polarized light with cold atoms inside an optical cavity \cite{JOSSE03,JOSSE04}.

It has been recently demonstrated that the single passage of a linearly polarized pump beam through a few-cm-long atomic vapor cell results in squeezing of the polarization orthogonal to that of the pump (vacuum squeezing) \cite{RIES03,MIKHAILOV08,MIKHAILOV09,AGHA10} as a consequence of the nonlinear optics mechanism known as polarization self-rotation (PSR) \cite{BOYDBOOK08,ROCHESTER01,MATSKO02}. Vacuum squeezing via PSR has been observed for the D1 \cite{MIKHAILOV08,MIKHAILOV09,AGHA10} and D2 \cite{RIES03} transitions using $^{87}$Rb vapor.  As noted in \cite{JOSSE03}, the existence of polarization squeezing can be inferred from these results.

In this article we present a study of polarization squeezing of light after traversing a quasi-resonant atomic medium. We study the fluctuations of a given stokes operator as a function of laser detuning and optical power and determine the set of orthogonal elliptical polarization pairs for which the corresponding Stokes operator is squeezed.

Our experiments are closely related to previous observations of squeezing via PSR \cite{RIES03,MIKHAILOV08,AGHA10}. In order to detect the quadrature dependent squeezing, these experiments utilize a homodyne detection setup in which the field been analyzed is made to interfere with a local oscillator. The transmitted pump field is conveniently used as the local oscillator. In order to vary the phase difference  $\theta$  between the vacuum field and the local oscillator these experiments use a Mach-Zender interferometer with controlled path difference between the two arms. Crucial to this technique is the quality of the wavefront overlap between the local oscillator and the analyzed field. In our work, we have chosen to control the phase difference  $\theta$ with the use of a retarder waveplate. This method has the advantage of introducing a stable phase angle (insensitive, in practice, to thermal drift or mechanical fluctuations). In addition, the wavefront overlap between the two polarization components is perfect. After this retarder, a second half-wave plate in combination with a polarizing beamsplitter (PBS) are used to split equally the light intensity towards two detectors. In this scheme, the light arriving at each detector corresponds to the two orthogonal elliptical polarization components $\mathbf{u_{\pm}}=\frac{1}{\sqrt{2}}(\mathbf{x}\pm e^{i\theta}\mathbf{y})$ of the beam after the atomic medium. The observed photocurrent difference signal measures a generalized Stokes operator corresponding to $S(\theta)\equiv n_{u_{+}(\theta)}-n_{u_{-}(\theta)}$ where $n_{u_{\pm}}$ are photon number operators. $S_{2}$ and $S_{3}$ correspond to $\theta=0$ and $\theta=\pi/2$ respectively.

\begin{figure}
\includegraphics[width=8.6cm]{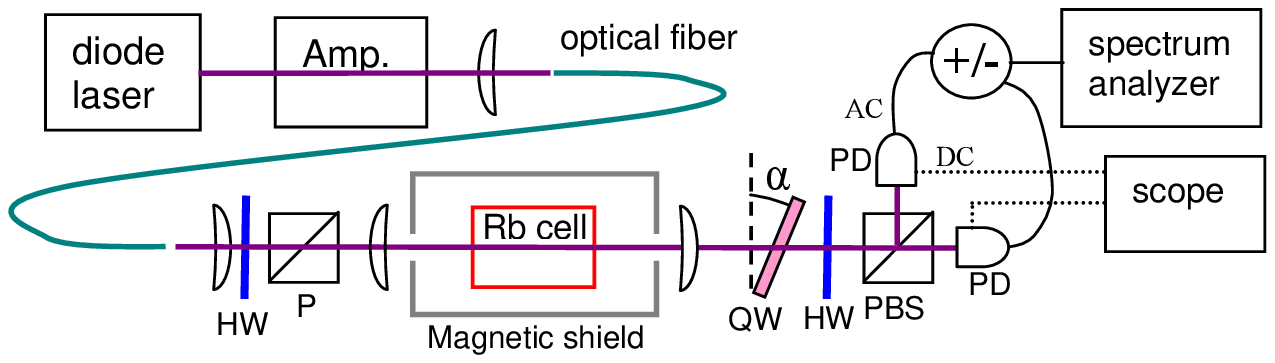}
\caption{\label{setup} (Color online) Experimental setup. P: polarizer, PBS: Polarizer beam splitter, HW: halfwave plate, QW: quarterwave plate, PD: photodiode.}
\end{figure}

We have studied polarization squeezing in light traversing a $^{87}$Rb vapor cell under experimental conditions very similar to those used by Mikhailov et al \cite{MIKHAILOV09}. The experimental setup is shown in Fig. \ref{setup}. The laser light is generated by an extended cavity diode laser followed by a tapered amplifier diode. The laser beam is spatially filtered with a single mode optical fiber. Up to 100 mW of laser light tunable to the D1 transitions of $^{87}$Rb (795 nm) are available after the fiber. The light beam is linearly polarized in the horizontal plane with a Glan polarizer. A half-wave plate, placed before the polarizer, is used to control the laser power at the atomic sample. The laser beam is focussed with a 30 cm lens near the center of a 5 cm long cylindrical glass cell containing highly isotopically enriched $^{87}$Rb vapor. The beam waist measured at $e^{-1}$ of the maximum intensity is 200 $\mu m$. The cell, placed inside a three-layer magnetic shield, is surrounded by a flexible silicon tube through which hot water is circulated to vary the vapor temperature and the atomic density accordingly. After the cell, the beam is recollimated and directed to the homodyne detection setup. High quantum efficiency ($\sim 90\%$) photodiodes are used for detection. The noise power of the amplified subtracted photocurrents is observed with a spectrum analyzer.

To vary the phase difference between horizontal and vertical polarization components, we have used a commercial zero order quarterwave plate (Thorlabs WPQ05M-780) with its slow axis oriented parallel to the polarization of the pump field. If the wave plate is perpendicular to the light beam, a phase angle $\theta = \pi/2$ is obtained. Different values of $\theta$ can be achieved by rotating the quarterwave plate a small angle $\alpha$ around its fast axis ($\alpha < 20^{\circ}$). The phase angle $\theta$ corresponding to a given value of $\alpha$ is calibrated by observing the interference pattern of the intensity transmitted though a linear polarizer rotated by $45^{\circ}$ with respect to the axis of the waveplate. It was verified that $\theta$ has no observable dependence on the laser frequency in the considered spectral range.

The determination of the noise SQL is obtained through the following procedure: Initially, the SQL noise level (shot noise) corresponding to a given DC light signal at the photodetectors is calibrated. For this, in order to keep the optical alignment unmodified, we detune the laser away from the atomic transitions and cool the atomic sample to room temperature so that its influence on light fluctuations becomes negligible. We have verified the characteristic linear dependence of the shot noise level on light power over all the considered power range. We then increase the vapor density by raising the cell temperature and analyze the light fluctuations while recording, at the same time, the average (equal) light level at the detectors. As the laser frequency is scanned, small systematic variations with laser frequency of the  incident light power ($< 10\%$) occur due to the sensitive alignment of the optical amplifier (due to saturation, the absorption by the atomic sample is negligible). Using the previous calibration, we determine the shot noise level for every value of the laser frequency.

\begin{figure}
\includegraphics[width=8.6cm]{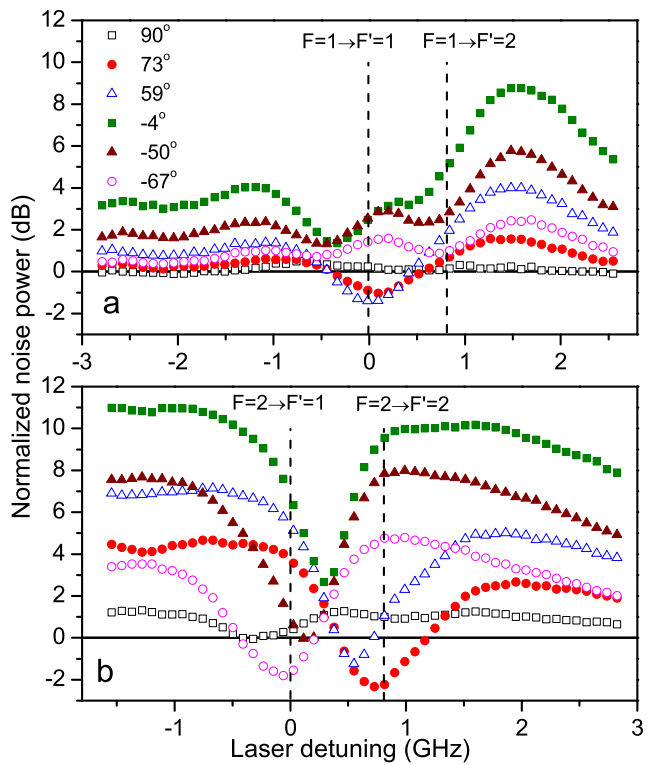}
\caption{\label{ruidoxdetuning} (Color online) Noise power at 2.7 MHz (100 KHz resolution bandwidth) as a function of laser detuning for different dephasing angles $\theta$. a)  $F=1\rightarrow F'$ transitions. b) $F=2\rightarrow F'$ transitions. Hollow squares: $\theta = 90\pm 1^{\circ}$, solid circles: $\theta = 73\pm 1^{\circ}$, hollow triangles: $\theta = 59\pm 2^{\circ}$, solid squares: $\theta = -4\pm 2^{\circ}$, solid triangles: $\theta = -50\pm 4^{\circ}$, hollow circles: $\theta = -67\pm 5^{\circ}$. Laser power: $P= 24$ mW, cell temperature: $T=73$ C.}
\end{figure}

The relative intensity noise, normalized to the shot noise level, for different values of the dephasing angle $\theta$ is presented in Fig. \ref{ruidoxdetuning} as a function of the laser frequency around the $F=1\rightarrow F'=1,2$ and $F=2\rightarrow F'=1,2$ transitions. The results are consistent with previous observations \cite{MIKHAILOV08,AGHA10} where only the minimum and maximum noise levels at a given laser frequency were reported. The largest squeezing occurs around the $F=2\rightarrow F'=1,2$ transitions. Significant squeezing is also observed around the $F=1\rightarrow F'=1$ transition. Consistently with previous observations and theoretical predictions, \cite{ROCHESTER01,MIKHAILOV09}, squeezing is not observed around the $F=1\rightarrow F'=2$ transition. Notice that the quadrature angles corresponding to squeezing are different for the three transitions. Squeezing is visible in the frequency range from 1 to 10 MHz which is the limit of our detection bandwidth.

\begin{figure}
\includegraphics[width=8.6cm]{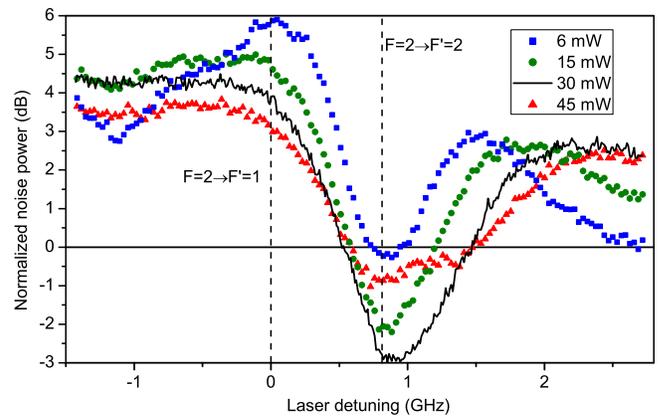}
\caption{\label{ruidoxpotencia} (Color online) Noise power at 2.7 MHz (resolution bandwidth: 100kHz, video bandwidht: 3 kHz) as a function of laser detuning around the $F=2\rightarrow F'$ transitions for different light powers. Squares: $P=6$ mW, circles:  $P=15$ mW, solid line: $P=30$ mW, triangles: $P=45$ mW. ($\theta = 73\pm 2^{\circ}, T=73$ C) }
\end{figure}

Fig. \ref{ruidoxpotencia} shows the recorded noise (at 2.7 MHz) for fixed $\theta = 73\pm 2^{\circ}$ as a function of laser detuning around the $F=2\rightarrow F'$  transitions for different values of the light power $P$. Squeezing is visible for $P$ larger than 6 mW. As the laser power is increased the spectral features are broadened and the position of the noise minimum is shifted toward larger laser frequencies \cite{AGHA10}.  The maximum  squeezing (solid trace in Fig. \ref{ruidoxpotencia}) was  $-2.9\pm 0.1$ dB. If we take into account the measured transmission losses in our setup (16\%) and the detector losses (10\%) the inferred squeezing is  $-4.7 \pm 0.2$ dB. It was observed  using 30 mW light power with a cell heated to 73 C corresponding to an atomic density $\sim 9\times 10^{11}$ cm $^{-3}$ \cite{ALCOCK84}. For light powers larger than 40 mW we observe that the amount of squeezing significantly decreases although Agha et al. \cite{AGHA10} were able to observe squeezing for light powers up to 200 mW. This difference could be due to the onset of saturation in the electronic amplifier used in our experiment since, in our detection scheme, the local oscillator cannot be independently attenuated.

We have extensively varied the parameters determining the squeezing generation efficiency.  Among these, the atomic density and light power (together with the light focussing geometry) have the stronger influence. We were able to observe squeezing for cell temperatures (atomic densities) ranging between 55 C $(2\times 10^{11}$ cm$^{-3})$ and 75 C $(1\times 10^{12}$ cm$^{-3})$.  We notice that the largest squeezing was obtained for conditions similar to those identified in the systematic parameter space survey reported in \cite{MIKHAILOV09}. The obtention of -3 dB squeezing is of special signification since this is the minimum squeezing level required for the implementation of several quantum information protocols such as e-cloning \cite{WEEDBROOK08}, entanglement swapping \cite{TAN99} and teleportation of coherent states \cite{GROSSHANS01}.

Our results can be analyzed from the point of view of squeezing via PSR \cite{MATSKO02,LEZAMA08,MIKHAILOV09}. However, the buildup of squeezing between two orthogonal polarization components of a linearly polarized beam, as it traverses an atomic sample, introduces a different and complementary perspective of the light-atom interaction process.  For a suitable choice of the atomic state basis, orthogonal light polarization components can be seen as coupling the two arms of a $\Lambda$ system formed by two degenerate ground state sublevels and an excited state. This configuration forms a typical electromagnetically induced transparency (EIT) scheme for which the buildup of classical as well as quantum correlations between the fields in a $\Lambda$ system have been predicted \cite{HARRIS93,AGARWAL93,JAIN94,GARRIDO03,HUSS04,FLEISCHHAUER05}. However, our actual atomic response, involving several hyperfine and Zeeman sublevels, is more complex than that of a three level $\Lambda$ system. It has been shown that PSR \cite{ROCHESTER01} and squeezing via PSR \cite{MIKHAILOV09} cannot be accurately described without the consideration of the complete excited state hyperfine structure. As suggested in \cite{MIKHAILOV09}, a four level scheme in a double $\Lambda$ or $X$ \cite{JOSSE04,HSU06} configuration is possibly the simplest model system that can explain the observed results. Further theoretical investigation on correlated field propagation in these systems is desirable \cite{CERBONESCHI96}.

In summary, we have studied the quantum correlations of different Stokes operators of a light beam after traversing a nearly resonant atomic vapor. Our observations show that quantum correlations naturally arise, under suitable experimental conditions, between two orthogonal elliptic polarizations of a linearly polarized light beam. The achieved squeezing is unprecedentedly large compared to previous observations of single-frequency continuous-wave squeezing generated in atomic samples and reaches the necessary level for the implementation of several quantum information protocols. The extreme simplicity of our setup based on standard polarization components, makes it particularly suitable for applications.

\begin{acknowledgments}
This work was supported by ANII, CSIC and PEDECIBA (Uruguayan agencies).
\end{acknowledgments}


\end{document}